\documentclass[showpacs,twocolumn,preprintnumbers,amsmath,amssymb]{revtex4}
\usepackage{graphicx}
\usepackage{dcolumn}
\begin{document}

\title{Output radiation from a degenerate parametric oscillator}

\author{Sintayehu Tesfa}
 \email{sint_tesfa@yahoo.com}
\affiliation{Physics Department, Addis Ababa University, P. O. Box 1176, Addis Ababa, Ethiopia}%


\date{\today}

\begin{abstract}We study the squeezing as well as the statistical properties
of the output radiation from a degenerate parametric oscillator
coupled to a squeezed vacuum reservoir employing the stochastic
differential equations associated with the normal ordering. It is
found that the degree of squeezing of the output radiation is less
than the corresponding cavity radiation. However, for output
radiation the correlation of the quadrature operators evaluated at
different times also exhibits squeezing, which is the reason for
quenching of the overall noise in one of the quadrature components
of the squeezing spectrum even when the oscillator is coupled to a
vacuum reservoir. Moreover, coupling the oscillator to the
squeezed vacuum reservoir enhances the squeezing exponentially and
it also increases the mean photon number.
\end{abstract}

 \maketitle

\section{INTRODUCTION}

Optical degenerate parametric oscillator is one of the most
interesting and well studied devices in the nonlinear quantum
optics \cite{1,2,3,4,5,6,7}. In a degenerate parametric
oscillator, a pump photon of frequency $2\omega$ is down converted
by nonlinear crystal into a pair of signal photons each of
frequency $\omega$.  Due to this inherent two-photon nature of the
interaction, the parametric oscillator is found to be a good
source of a squeezed light. The maximum degree of squeezing of the
cavity radiation coupled to an ordinary vacuum is found to be
$50\%$ at the critical point by many authors using various
approaches \cite{1,2,3}. This limitation in squeezing is
associated with the leakage through the mirror and the inevitable
amplification of the quantum fluctuations in the cavity due to
thermal heating. Since the vacuum field has no definite phase,
squeezing of the cavity radiation when coupled to a vacuum
reservoir would be degraded. However, if the ordinary vacuum is
replaced by a squeezed vacuum reservoir, it is expected that the
fluctuations entering the cavity is biased and hence the squeezing
of the cavity radiation could be enhanced \cite{2,3,4,8} provided
that the reservoir is squeezed in the right quadrature.

Theoretical analysis of the quantum fluctuations and photon
statistics of the cavity radiation of a degenerate parametric
oscillator coupled to a squeezed vacuum reservoir has been
exhaustively made by many authors \cite{1,3,4}, although it is the
output radiation which is practically accessible for an
experiment. To our knowledge, there is no a thorough study of the
squeezing and statistical properties of the output radiation
except for some discussions related with the squeezing spectrum
\cite{1,3,5}. For instance,  Collett and Gardiner \cite{5}
claimed, after comparing the variance of the quadrature operators
for the cavity radiation with the squeezing spectrum for the
output radiation, that the squeezing of the output radiation is
much better than the cavity radiation. Since the output squeezing
spectrum corresponds to the correlation of the output quadrature
operators evaluated at different times, $t$ and $t+\tau$, in the
frequency domain, we believe that it would not be right comparing
the squeezing spectrum with the variance of the cavity radiation.

We hence devoted this work to the analysis of the output radiation
of a degenerate parametric oscillator coupled to a squeezed vacuum
reservoir using the input-output relation introduced by Gardiner
and Collett \cite{9}. We device a means of evaluating a
correlation between the noise forces arising from the reservoir
and vacuum fluctuations in the cavity that required in studying
the properties of the  radiation outside the cavity. Moreover, we
followed an adhoc approach in casting the characteristic function
describing the cavity radiation onto the corresponding output
variables. On the basis of these approaches we study the squeezing
as well as the statistical properties of the output radiation
employing the stochastic differential equations associated with
the normal ordering. In particular, we calculate the quadrature
variance, squeezing spectrum, photon number distribution, and
power spectrum for the output radiation. We also compare the
results we obtained with the corresponding values for the cavity
radiation. We believe that the approach followed in this work
could also be used to study the output radiation of other quantum
optical systems.

\section{ STOCHASTIC DIFFERENTIAL EQUATIONS}

The degenerate parametric oscillator can be described in the
interaction picture and in the rotating-wave approximation, upon
treating the pump radiation classically, by the Hamiltonian of the
form
\begin{align}\label{dp01}\hat{H}={i\varepsilon\over2}\big(\hat{a}^{\dagger^{2}}
-\hat{a}^{2}\big),\end{align} where $\varepsilon$  is proportional
to the amplitude of the external coherent radiation which is taken
to be a real-positive constant and $\hat{a}$ is the annihilation
operator for the signal mode. We consider the case in which a
continuum mode of squeezed vacuum centered at frequency $\omega$
is allowed to enter the cavity through one of the coupler mirrors.
In this case, the master equation describing the degenerate
parametric oscillator coupled to a broadband squeezed vacuum
reservoir in the interaction picture is found using the standard
procedure \cite{10} to be
\begin{align}\label{dp02}{d\hat{\rho}\over dt}& =
{\varepsilon\over2}\left(
\hat{a}^{\dagger^{2}}\hat{\rho}-\hat{a}^{2}\hat{\rho}-\hat{\rho}\hat{a}^{\dagger^{2}}+
\hat{\rho}\hat{a}^{2}\right)\notag\\&
+{\kappa(N+1)\over2}\left(2\hat{a}\hat{\rho}\hat{a}^{\dagger}-\hat{a}^{\dagger}\hat{a}\hat{\rho}
- \hat{\rho}\hat{a}^{\dagger}\hat{a}\right) \notag\\&+{\kappa
N\over2}\left(2\hat{a}^{\dagger}\hat{\rho}\hat{a}-\hat{a}\hat{a}^{\dagger}\hat{\rho}
- \hat{\rho}\hat{a}\hat{a}^{\dagger}\right) \notag\\&+ {\kappa
M\over2}\left(\hat{a}^{2}\hat{\rho}-2\hat{a}\hat{\rho}\hat{a} +
\hat{\rho}\hat{a}^{2}-2\hat{a}^{\dagger}\hat{\rho}\hat{a}^{\dagger}+
\hat{a}^{\dagger^{2}}\hat{\rho} +
\hat{\rho}\hat{a}^{\dagger^{2}}\right),\end{align} where $N$ and
$M$ represent the squeezed vacuum reservoir, with
$N=\sinh^{2}(r),$
 $M=\sinh(r)\cosh(r)$, $\kappa$ is the cavity damping constant, and $r$ is the squeeze parameter.

We next proceed to obtain the stochastic differential equations
pertinent to the cavity mode variables using the master equation
\eqref{dp02}. To this end, employing the fact that
\begin{align}\label{dp03}{d\over dt}\langle\hat{a}(t)\rangle = Tr\left({d\hat{\rho}
\over dt}\hat{a}\right),\end{align} it is possible see that
\begin{align}\label{dp04}{d\over dt}\langle\hat{a}(t)\rangle
=-{\kappa\over2}\langle\hat{a}(t)\rangle+\varepsilon\langle\hat{a}^{\dagger}(t)\rangle,\end{align}
\begin{align}\label{dp05}{d\over
dt}\langle\hat{a}^{\dagger}(t)\hat{a}(t)\rangle&
=-\kappa\langle\hat{a}^{\dagger}(t)\hat{a}(t)\rangle\notag\\&+\varepsilon[
\langle\hat{a}^{2}(t)\rangle+\langle\hat{a}^{\dagger^{2}}(t)\rangle]
+\kappa N,\end{align}
\begin{align}\label{dp06}{d\over dt}\langle\hat{a}^{2}(t)\rangle
=-\kappa\langle\hat{a}^{2}(t)\rangle+2\varepsilon\langle\hat{a}^{\dagger}(t)\hat{a}(t)\rangle
+\varepsilon+\kappa M.\end{align} We notice that the operators in
Eqs. \eqref{dp04}, \eqref{dp05}, and \eqref{dp06} are in the
normal order. Hence the corresponding expressions in the c-number
variables associated with the normal ordering take the form
\begin{align}\label{dp07}{d\over dt}\langle\alpha(t)\rangle
=-{\kappa\over2}\langle\alpha(t)\rangle+\varepsilon\langle\alpha^{*}(t)\rangle,\end{align}
\begin{align}\label{dp08}{d\over
dt}\langle\alpha^{*}(t)\alpha(t)\rangle&
=-\kappa\langle\alpha^{*}(t)\alpha(t)\rangle\notag\\&+\varepsilon[
\langle\alpha^{2}(t)\rangle+\langle\alpha^{*^{2}}(t)\rangle]
+\kappa N,\end{align}
\begin{align}\label{dp09}{d\over dt}\langle\alpha^{2}(t)\rangle
=-\kappa\langle\alpha^{2}(t)\rangle+2\varepsilon\langle\alpha^{*}(t)\alpha(t)\rangle+\varepsilon+\kappa
M.\end{align} On the basis of Eq. \eqref{dp07}, one can write that
\begin{align}\label{dp10}{d\over dt}\alpha(t)
=-{\kappa\over2}\alpha(t)+\varepsilon\alpha^{*}(t) +
F(t),\end{align} where $F(t)$ is a complex Gaussian white noise
force the mean and correlation properties of which remain to be
determined. It is not difficult to see that Eq. \eqref{dp07} will
be equal to the expectation value of Eq. \eqref{dp10} provided
that
\begin{align}\label{dp11}\langle F(t)\rangle=0.\end{align}
On the other hand, one can verify applying Eqs. \eqref{dp07},
\eqref{dp08}, \eqref{dp09}, and \eqref{dp10} along with the fact
that the noise force at time $t$ does not correlate with the
cavity mode variables at earlier times that
\begin{align}\label{dp12}\langle F(t')F(t)\rangle
=(\kappa M+\varepsilon)\delta(t-t'),\end{align}
\begin{align}\label{dp13}\langle F^{*}(t)F(t)\rangle=\kappa N\delta(t-t').\end{align}
We observe that Eqs. \eqref{dp11}, \eqref{dp12}, and \eqref{dp13}
represent the mean and correlation properties of the noise force.

In general, it is possible to express Eq. \eqref{dp10} in a more
convenient from, upon introducing two variables defined by
\begin{align}\label{dp14}\alpha_{\pm}(t) = \alpha^{*}(t) \pm
\alpha(t),\end{align} as
\begin{align}\label{dp15}\frac{d}{dt}\alpha_{\pm} =
-\frac{1}{2}\lambda_{\pm}\alpha_{\pm} + F_{\pm}(t),\end{align}
with
\begin{align}\label{dp16}\lambda_{\pm} = \kappa \mp
2\varepsilon,\end{align}
\begin{align}\label{dp17}F_{\pm}(t)=F^{*}(t)\pm F(t),\end{align}
where the formal solution of Eq. \eqref{dp15} can be written as
\begin{align}\label{dp18}\alpha_{\pm}(t+\tau) &=
\alpha_{\pm}(t)e^{-\frac{\lambda_{\pm}}{2}\tau}\notag\\& +
\int_{0}^{\tau}e^{-\frac{\lambda_{\pm}}{2}(\tau -
\tau')}F_{\pm}(t+\tau')d\tau'.\end{align} Now in view of Eqs.
\eqref{dp14} and \eqref{dp18} into account, we get
\begin{align}\label{dp19}\alpha(t+\tau)& = A_{+}(\tau)\alpha(t) + A_{-}(\tau)\alpha^{*}(t)
\notag\\&+ B_{+}(t+\tau) - B_{-}(t+\tau),\end{align} in which
\begin{align}\label{dp20}A_{\pm}(\tau) = \frac{1}{2}\big[e^{-\frac{\lambda_{+}}{2}\tau}
\pm e^{-\frac{\lambda_{-}}{2}\tau}\big],\end{align}
\begin{align}\label{dp21}B_{\pm}(t+\tau) = \frac{1}{2}\int_{0}^{\tau}e^{-\frac{\lambda_{\pm}}{2}(\tau - \tau')}
F_{\pm}(t+\tau')d\tau'.\end{align} Next upon setting $t=0$ in Eq.
\eqref{dp19} and then replacing $\tau$ by $t$, we have
\begin{align}\label{dp22}\alpha(t)& = A_{+}(t)\alpha(0) + A_{-}(t)\alpha^{*}(0) +
B_{+}(t) - B_{-}(t).\end{align} It perhaps worth mentioning that
Eqs. \eqref{dp19} and \eqref{dp22} is applied in calculating
various quantities of interest. We also realize that these
solutions would be well-behaved functions at steady state, if
$\kappa>2\varepsilon$. Hence we designate $\kappa=2\varepsilon$ as
a critical point.

\section{ QUADRATURE FLUCTUATIONS}

The squeezing properties of a single-mode output radiation can be
described with the aid of the quadrature operators defined by
\begin{align}\label{dp23}\hat{a}^{out}_{+}=\hat{a}_{out}^{\dagger}+\hat{a}_{out}\end{align}
and
\begin{align}\label{dp24}\hat{a}^{out}_{-}=i(\hat{a}_{out}^{\dagger}-\hat{a}_{out}).\end{align}

\subsection{Quadrature variance}

The variances of the quadrature operators \eqref{dp23} and
\eqref{dp24} can be put, using the boson commutation relations for
the output radiation, in the form
\begin{align}\label{dp25}\Delta a^{2}_{\pm(out)}&=1\pm\big[\langle\hat{a}_{out}^{\dagger^{2}}(t)\rangle
+\langle\hat{a}_{out}^{2}(t)\rangle
\pm2\langle\hat{a}_{out}^{\dagger}(t)\hat{a}_{out}(t)\rangle
\notag\\&+\langle\hat{a}_{out}^{\dagger}(t)\rangle^{2}+\langle\hat{a}_{out}(t)\rangle^{2}\pm2\langle\hat{a}_{out}^{\dagger}
(t)\rangle\langle\hat{a}_{out}(t)\rangle\big].\end{align} We
notice that the operators in Eq. \eqref{dp25} are in the normal
order. Hence the corresponding expression in terms of the c-number
variables associated with the normal ordering would be
\begin{align}\label{dp26}\Delta a^{2}_{\pm(out)}&=1\pm\big[\langle\alpha_{out}^{*^{2}}(t)\rangle
+\langle\alpha_{out}^{2}(t)\rangle
\pm2\langle\alpha_{out}^{*}(t)\alpha_{out}(t)\rangle
\notag\\&+\langle\alpha_{out}^{*}(t)\rangle^{2}+\langle\alpha_{out}(t)\rangle^{2}\pm2\langle\alpha_{out}^{*}
(t)\rangle\langle\alpha_{out}(t)\rangle\big].\end{align} It is a
well established fact that the output radiation can be defined in
terms of the cavity mode variables \cite{9} as
\begin{align}\label{dp27}\alpha_{out}(t)=\sqrt{\kappa}\alpha(t)-
{1\over\sqrt{\kappa}}F_{R}(t),\end{align} where $F_{R}(t)$ is the
noise force associated with the reservoir modes and satisfies the
correlations:
\begin{align}\label{dp28}\langle F_{R}(t)\rangle =0,\end{align}
\begin{align}\label{dp29}\langle F_{R}^{*}(t)F_{R}(t')\rangle =\kappa N\delta(t-t'),\end{align}
\begin{align}\label{dp30}\langle F_{R}(t)F_{R}(t)\rangle =\kappa M\delta(t-t').\end{align}
If the cavity mode is taken to be initially in a vacuum state, Eq.
\eqref{dp26} reduces to
\begin{align}\label{dp31}\Delta
a^{2}_{\pm(out)}&=1+2\langle\alpha_{out}^{*}(t)\alpha_{out}(t)\rangle\notag\\&\pm\big[
\langle\alpha_{out}^{*^{2}}(t)\rangle+\langle\alpha_{out}^{2}(t)\rangle\big].\end{align}

We next seek to determine the correlations involved in Eq.
\eqref{dp31}. To this end, making use of Eq. \eqref{dp27}, we see
that
\begin{align}\label{dp32}\langle\alpha_{out}^{*}(t)\alpha_{out}(t)\rangle
& =\kappa\langle\alpha^{*}(t)\alpha(t)\rangle
+{1\over\kappa}\langle F^{*}_{R}(t)F_{R}(t)\rangle \notag\\&-
\langle\alpha^{*}(t)F_{R}(t)\rangle -\langle
F_{R}^{*}(t)\alpha(t)\rangle.\end{align} We observe that the noise
force $F(t)$ represents the contribution from the cavity vacuum
fluctuations and reservoir, and hence can be expressed
\begin{align}\label{dp33}F(t)=F_{C}(t)+F_{R}(t),\end{align}
where $F_{C}(t)$ is the noise force corresponding to the system in
the cavity in the absence of damping. Since the noise force for
the reservoir $F_{R}(t)$ does not correlate with $F_{C}(t)$ and
the system variables at the earlier times, it is not difficult to
verify that
\begin{align}\label{dp34}\langle F_{R}(t)\alpha^{*}(t)\rangle +
\langle F^{*}_{R}(t)\alpha(t)\rangle = \kappa N.\end{align} Thus
on account of Eqs. \eqref{dp29}, \eqref{dp32}, and \eqref{dp34},
we get
\begin{align}\label{dp35}\langle \alpha_{out}^{*}(t)\alpha_{out}(t)\rangle =
\kappa\langle \alpha^{*}(t)\alpha(t)\rangle
+N(1-\kappa).\end{align} We are usually interested in steady state
values in quantum optics. In this regard, it is easy to check
applying Eqs. \eqref{dp08} and \eqref{dp09} that
\begin{align}\label{dp36}\langle
\alpha^{*}(t)\alpha(t)\rangle_{ss}
={2\varepsilon^{2}\over\kappa^{2}-4\varepsilon^{2}}
+{\kappa(N\kappa+2\varepsilon
M)\over\kappa^{2}-4\varepsilon^{2}},\end{align}
 as a result the
mean photon number for the output radiation at steady state takes
the form
\begin{align}\label{dp37}\langle
\alpha_{out}^{*}(t)\alpha_{out}(t)\rangle_{ss}&
=N+{2\kappa\varepsilon^{2}\over\kappa^{2}-4\varepsilon^{2}}
+{2\varepsilon\kappa(2N\varepsilon+\kappa
M)\over\kappa^{2}-4\varepsilon^{2}}.\end{align}
\begin{center}
\begin{figure}[hbt]
\centerline{\includegraphics [height=5.5cm,angle=0]{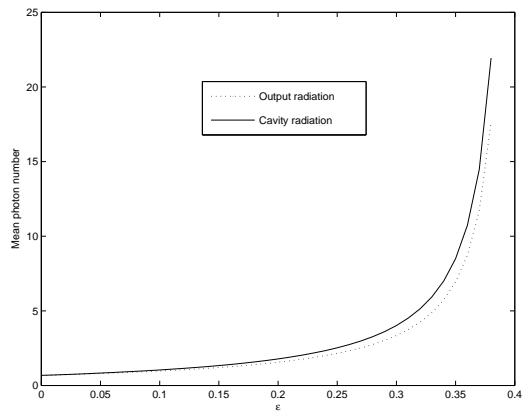} }
\caption{Plots of the mean photon number of the output and cavity
radiation at steady state for $\kappa =0.8$ and $r=0.75$.}
\end{figure}
\end{center}

\begin{center}
\begin{figure}[hbt]
\centerline{\includegraphics [height=5.5cm,angle=0]{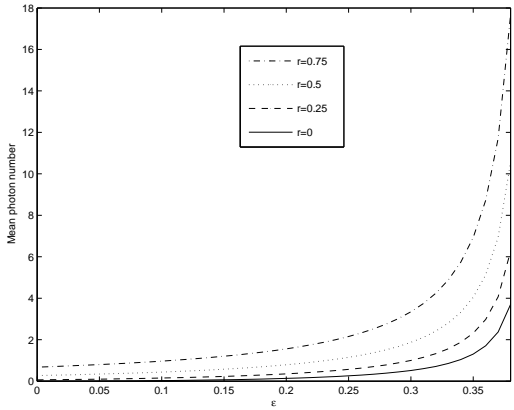} }
\caption{Plots of the mean photon number of the output radiation
at steady state for $\kappa =0.8$ and different values of $r$.}
\end{figure}
\end{center}

\noindent We notice that the first term in Eq. \eqref{dp37} is the
contribution from the squeezed vacuum radiation that exists
outside the cavity, the second term corresponds to the light
produced by the parametric oscillator and then escapes through the
mirror, whereas the last term represents the correlation between
the oscillator and reservoir due to the associated noises. We
realize that the larger the intensity of the external coherent
radiation, the more probable it is converted into the cavity
radiation by the crystal. It is a known fact that the squeezed
parameter is directly related to the mean photon number of the
squeezed vacuum modes. Thus the more the reservoir modes are
reflected from the mirror, the more the mean photon number outside
the cavity would be. We hence clearly see from Fig. 2 that the
mean photon number for the output radiation increases with the
amplitude of the coherent radiation and squeeze parameter of the
reservoir. However, in spite of the presence of the squeezed
vacuum radiation outside the cavity, we clearly see from Fig. 1
that the mean photon number is greater in the cavity, since only
few of the photons in the cavity can escape through the mirror.

Furthermore, it is  possible to verify following  a similar
procedure that
\begin{align}\label{dp38}\langle \alpha_{out}^{2}(t)\rangle_{ss} =
M+{\kappa^{2}\varepsilon\over\kappa^{2}-4\varepsilon^{2}}
+{2\kappa\varepsilon(\kappa N+2\varepsilon
M)\over\kappa^{2}-4\varepsilon^{2}},\end{align} so that the
variance of the output quadrature operators are found to be
\begin{align}\label{dp39}\Delta a_{\pm(out)}^{2}=e^{\pm2r}\left[1\pm
{2\varepsilon\kappa\over\kappa\mp2\varepsilon}\right],\end{align}
which reduces at critical point to
\begin{align}\label{dp40a}\Delta
a_{+(out)}^{2}=\infty,\end{align}
\begin{align}\label{dp40}\Delta
a_{-(out)}^{2}=e^{-2r}\left[1- 0.5k\right].\end{align} In
addition, the quadrature variances for the cavity radiation turn
out following a similar approach to be
\begin{align}\label{dp41}\Delta
a^{2}_{\pm}=\left({\kappa\over\kappa\mp2\varepsilon}\right)e^{\pm2r},\end{align}
in which at critical point $\Delta a^{2}_{+}\rightarrow\infty$ and
$\Delta a^{2}_{-}={1\over2}e^{\pm2r}$.

\begin{center}
\begin{figure}[hbt]
\centerline{\includegraphics [height=5.5cm,angle=0]{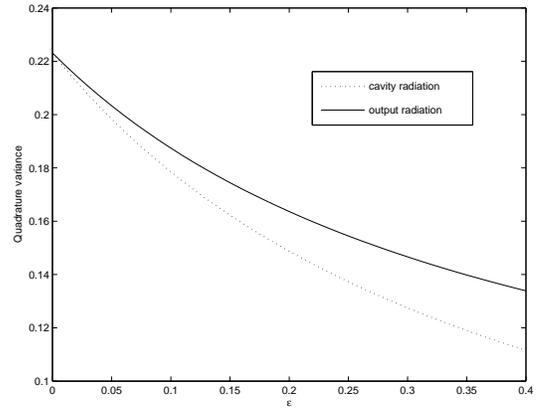}}
\caption {Plots of the quadrature variance for the output and
cavity radiation at steady state for $\kappa =0.8$ and
 $r=0.75$.}
\end{figure}
\end{center}
Since $\kappa<1$,  one can easily infer from Eq. \eqref{dp40} that
the squeezing of the output radiation is less than 50\% for a
vacuum reservoir at critical point. We also see from Fig. (3) that
the degree of squeezing of the cavity radiation is far better than
the output radiation except for very small values of the amplitude
of the driving radiation. The unbiased noise fluctuations arising
while the radiation crosses the boundary (coupler mirror) is
believed to be the reason for the reduction of the degree of
squeezing for the output radiation.

\subsection{ Squeezing spectrum}

In this section. we seek to calculate the squeezing spectrum for
the output radiation, which can be defined as
\begin{align}\label{dp44}S^{out}_{\pm}(\omega)& = 2Re\int_{0}^{\infty} e^{i\omega\tau}
\langle\hat{a}^{out}_{\pm}(t+\tau),\;
\hat{a}^{out}_{\pm}(t)\rangle_{ss}d\tau,\end{align} where
\begin{align}\label{dp45}\langle\hat{a}^{out}_{\pm}(t+\tau),\;
\hat{a}^{out}_{\pm}(t)\rangle &
=\langle\hat{a}^{out}_{\pm}(t+\tau)\hat{a}^{out}_{\pm}(t)\rangle
\notag\\&-\langle\hat{a}^{out}_{\pm}(t+\tau)\rangle\langle
\hat{a}^{out}_{\pm}(t)\rangle.\end{align} With the application of
\eqref{dp14}, \eqref{dp23}, and \eqref{dp24}, Eq. \eqref{dp44} can
be put, for the cavity radiation initially in vacuum state, in
terms of the c-number variables associated with the normal
ordering as
\begin{align}\label{dp46}S^{out}_{\pm}(\omega) = 1\pm2Re\int_{0}^{\infty}
e^{i\omega\tau}
\langle\alpha^{out}_{\pm}(t+\tau)\alpha^{out}_{\pm}(t)\rangle_{ss}d\tau.\end{align}
Now making use  of Eq. \eqref{dp27}, we write
\begin{align}\label{dp47}\langle\alpha^{out}_{\pm}(t+\tau)\alpha^{out}_{\pm}(t)
\rangle_{ss}& =\kappa\langle\alpha_{\pm}(t+\tau)\alpha_{\pm}(t)
\rangle_{ss} \notag\\&+ {1\over\kappa}\langle
F_{R\pm}(t+\tau)F_{R\pm}(t)
\rangle_{ss}\notag\\&-\langle\alpha_{\pm}(t+\tau)F_{R\pm}(t)\rangle_{ss}\notag\\&-
\langle\alpha_{\pm}(t)F_{R\pm}(t+\tau) \rangle_{ss},\end{align} in
which
\begin{align}\label{dp48}F_{R\pm}(t)=F_{R}^{*}(t)\pm F_{R}(t).\end{align}
On the basis of the fact that the noise force does not correlate
with system variables at earlier times, it is possible to verify
that
\begin{align}\label{dp49}\langle\alpha_{\pm}(t+\tau)\alpha_{\pm}(t)\rangle_{ss}
= {2\over\kappa\mp2\varepsilon}[\kappa(M\pm
N)+\varepsilon]e^{-{\lambda_{\pm}\over2}\tau},\end{align}
\begin{align}\label{dp50}\langle F_{R\pm}(t)F_{R\pm}(t+\tau)\rangle
=2\kappa(M\pm N)\delta(\tau),\end{align}
\begin{align}\label{dp51}\langle\alpha_{\pm}(t+\tau)F_{R\pm}(t)\rangle
=2\kappa(M\pm N)e^{-{\lambda_{\pm}\over2}\tau},\end{align} as a
result employing Eqs. \eqref{dp46}, \eqref{dp47}, \eqref{dp49},
\eqref{dp50}, and \eqref{dp51}, we find
\begin{align}\label{dp52}S^{out}_{\pm}(\omega)=e^{\pm2r}\left[1\pm
{8\varepsilon\kappa\over(\kappa\mp2\varepsilon)^2+4\omega^{2}}\right].\end{align}
At critical point and $\omega=0$, we see that $S^{out}_{-}(0)=0$.
\begin{center}
\begin{figure}[hbt]
\centerline{\includegraphics [height=5cm,angle=0]{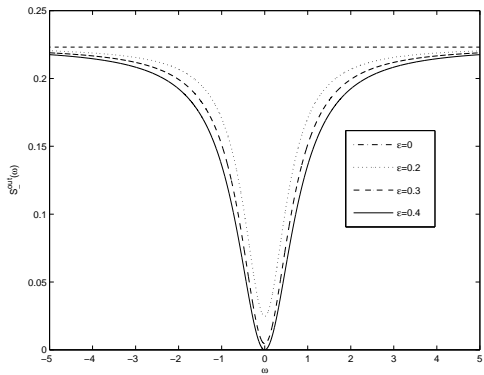}}
\caption {Plots of the squeezing spectrum of the output radiation
at steady state for $\kappa =0.8$, $r=0.75$, and different values
of $\varepsilon$.}
\end{figure}
\end{center}

We notice from Eq. \eqref{dp52} that for $\varepsilon=0$, the
squeezing spectrum is completely flat with its value depending on
the squeeze parameter. However, for $\varepsilon\ne0$ the depth of
the squeezing spectrum at $\omega=0$ increases with the amplitude
of the coherent radiation and it would be zero at critical point
irrespective of the values of the squeeze parameter. It is not
difficult to see that the width of the squeezing spectrum deceases
with the amplitude of the coherent radiation.  We realize upon
comparing the results shown in Figs. 3 and 4 that the squeezing
spectrum can take smaller values than the corresponding quadrature
variance of the output radiation for the same $r$ and
$\varepsilon$. But this does not mean that the squeezing of the
output radiation is greater than the cavity radiation, since the
squeezing spectrum corresponds to the correlation of the
quadrature operators evaluated at different times unlike the
quadrature variance of the cavity radiation which is associated
with the correlation of the quadrature variance at equal time.

 \section{PHOTON STATISTICS}

\subsection{Photon number distribution}

The probability for finding $n$ photons in a single-mode radiation
can be expressed in terms of the pertinent Q-function as
\begin{align}\label{dp54}P(n) =
\frac{\pi}{n!}\frac{\partial^{2n}}{\partial\alpha^{n}\partial\alpha^{*^{n}}}
\big[Q(\alpha)\exp(\alpha^{*}\alpha)\big]_{\alpha^{*} = \alpha
=0},\end{align} where the Q-function for the single-mode radiation
can be defined as
\begin{align}\label{dp55}Q(\alpha,t) = \frac{1}{\pi^{2}}\int
d^{2}z\;\phi(z,z,^{*},t)\exp[z^{*}\alpha -
z\alpha^{*}],\end{align} in which the antinormally ordered
characteristic function $\phi(z,z^{*},t)$ is given by
\begin{align}\label{dp56}\phi(z,z^{*},t) =
Tr\{\hat{\rho}(0)e^{-z^{*}\hat{a}(t)}e^{z\hat{a}^{\dagger}(t)}\}.\end{align}
Using the identity
\begin{align}\label{dp57}e^{\hat{A}}e^{\hat{B}} =
e^{\hat{B}}e^{\hat{A}}e^{[\hat{A},\;\hat{B}]},\end{align} Eq.
\eqref{dp56} can  be written  in terms of the c-number variables
associated with the normal ordering as
\begin{align}\label{dp58}\phi(z,z^{*},t) =
e^{-z^{*}z}\langle\exp(z\alpha^{*}(t)
-z^{*}\alpha(t))\rangle.\end{align} We note that $\alpha$ is a
Gaussian variable with zero mean. Hence one can readily verify
that \cite{11}
\begin{align}\label{dp59}\langle\exp(z\alpha^{*}(t)-z^{*}\alpha(t))\rangle =
\exp\left[\frac{1}{2}\langle(z\alpha^{*}(t) -z^{*}\alpha(t))^{2}
\rangle\right].\end{align} Therefore, it is possible to express
the antinormally ordered characteristic function for the output
radiation as
\begin{align}\label{dp60}\phi^{out}(z,z^{*},t) &= e^{-z^{*}z
}\exp\left[\frac{1}{2}\left(z^{2}\langle\alpha^{*^{2}}_{out}(t)\rangle
+z^{*^{2}}\langle\alpha^{2}_{out}(t)\rangle\right.\right.\notag\\&\left.\left.-
2z^{*}z\langle\alpha^{*}_{out}(t)\alpha_{out}(t)\rangle\right)\right],\end{align}
which can also be put at steady state in the form
\begin{align}\label{dp61}\phi^{out}(z,z^{*}) = \exp[- z^{*}za + {b\over2}(z^{2}
+z^{*^{2}})],\end{align} where $a=1+\bar{n}$ and $
b=\langle\alpha^{2}_{out}(t)\rangle_{ss}$, in which $\bar{n}$
represents the mean photon number for the output radiation at
steady state \eqref{dp37}. Hence substituting Eq. \eqref{dp61}
 into \eqref{dp55} and
then performing the integration, the Q-function describing the
output radiation is found to have the form
\begin{align}\label{dp62}Q^{out}(\alpha) =
\frac{\sqrt{u^{2}-v^{2}}}{\pi}\exp\left[ -u\alpha^{*}\alpha
+{v\over2}(\alpha^{2}+\alpha^{*^{2}})\right],\end{align} with $u =
\frac{a}{a^{2}-b^{2}}$ and $v = \frac{b}{a^{2}-b^{2}}$.

On the other hand, insertion of Eqs. \eqref{dp62} into
\eqref{dp54} results in
\begin{align}\label{dp63}P^{out}(n)&={\sqrt{u^{2}-v^{2}}\over n!}
{\partial^{2n}\over\partial\alpha^{*^{n}}
\partial\alpha^{n}}\notag\\&\times
\left[\exp\left[(1-u)\alpha^{*}\alpha+{v\over2}\big(\alpha^{*^{2}}
+\alpha^{2}\big) \right]\right]_{\alpha=\alpha^{*}=0},\end{align}
from which follows
\begin{align}\label{dp64}P^{out}(n)&={\sqrt{u^{2}-v^{2}}\over n!}
\left[\sum^{\infty}_{i,j,k=0}{(1-u)^{i}v^{k+j}\over2^{k+j}i!j!k!}\right.\notag\\&\left.\times
{(2k+i)!\over(2k+i-n)!}
{(2j+i)!\over(2j+i-n)!}\right.\notag\\&\left.\times
\alpha^{*^{2k+i-n}}\alpha^{2j+i-n}\right]_{\alpha=\alpha^{*}=0}.\end{align}
Now applying the condition, $ \alpha^{*}=\alpha=0$, we note that
Eq. \eqref{dp64} would be different from zero provided that
$2k+i-n=0$ and $2j+i-n=0$, as a result $k=j$ and $j={n-i\over2}$.
Consequently, the photon number distribution for the output
radiation finally takes the form
\begin{align}\label{dp65}P^{out}(n) &=
{n!\over\left[1+2\bar{n}+\bar{n}^{2}-\langle\alpha^{2}_{out}(t)
\rangle^{2}_{ss}\right]^{2n+1\over2}}\notag\\&\times
\sum_{i=0}^{n}{\left[\bar{n}^{2}+\bar{n}-\langle\alpha^{2}_{out}(t)\rangle_{ss}^{2}\right]
^{i}\langle\alpha^{2}_{out}(t)\rangle_{ss}^{n-i}\over
2^{n-i}i![({n-i\over2})!]^{2}}.\end{align} We notice that there is
a finite probability for counting odd number of photons outside
the cavity, even though degenerate parametric oscillator generates
pairs of photons. This could be related to the fact that odd
number of photons can escape through the coupler mirror.

\subsection{ Power spectrum}

The power spectrum for the output radiation can be expressed in
terms of the c-number variables associated with the normal
ordering as
\begin{align}\label{dp66}S^{out}(\omega) = 2Re\int_{0}^{\infty}
\langle\alpha_{out}^{*}(t)\alpha_{out}(t+\tau)\rangle_{ss}e^{i\omega\tau}.\end{align}
Here with the aid of Eqs. \eqref{dp12}, \eqref{dp13},
\eqref{dp19}, \eqref{dp20}, \eqref{dp21}, \eqref{dp27},
\eqref{dp29}, and \eqref{dp30}, we see that
\begin{align}\label{dp67}\langle\alpha^{*}_{out}(t)\alpha_{out}(t+\tau)\rangle_{ss}
&= {\kappa\over2}(\langle\alpha^{*}(t)\alpha(t)\rangle_{ss}
+\langle\alpha^{2}(t)\rangle_{ss}\notag\\&-(M+N))e^{-{\lambda_{+}\over2}\tau}
\notag\\&+{\kappa\over2}(\langle\alpha^{*}(t)\alpha(t)\rangle_{ss}
-\langle\alpha^{2}(t)\rangle_{ss}\notag\\&-(N-M))e^{-{\lambda_{-}\over2}\tau}
+ N\delta(\tau),\end{align} as a result the power spectrum
\eqref{dp66} for the output radiation turns out to be
\begin{align}\label{dp68}S^{out}(\omega) &= N +2\kappa\varepsilon\notag\\&\times\left[{1+2N+2M
\over(\kappa-2\varepsilon)^{2}+4\omega^{2}}+{1-2N+2M
\over(\kappa+2\varepsilon)^{2}+4\omega^{2}}\right].\end{align} In
a similar manner the power spectrum for the cavity radiation is
found to be
\begin{align}\label{dp69}S(\omega) &=
2\left[{\kappa(N+M)+\varepsilon
\over(\kappa-2\varepsilon)^{2}+4\omega^{2}}+{\kappa(N-M)-\varepsilon
\over(\kappa+2\varepsilon)^{2}+4\omega^{2}}\right].\end{align}

\begin{center}
\begin{figure}[hbt]
\centerline{\includegraphics [height=5.5cm,angle=0]{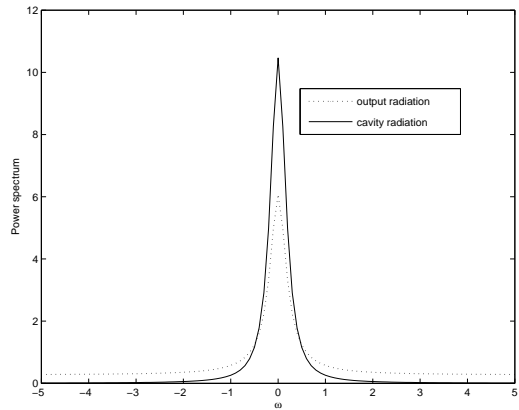}}
\caption {Plots of the power spectrum of the output and cavity
radiation at steady state for $\kappa =0.8$, $r=0.5$, and
$\varepsilon=0.2$.}
\end{figure}
\end{center}

\noindent We realize from Eqs. \eqref{dp68} and \eqref{dp69} that
the width of the cavity and output power spectra is independent of
the squeeze parameter, but it decreases with the amplitude of the
driving radiation. As clearly shown in Fig. 5 the power spectrum
for cavity radiation is more picked and narrowed. However, the
width for the cavity and output radiation is the same for the same
values of $\varepsilon$.

\section{ CONCLUSION}

In this paper, we present a thorough study of the squeezing and
statistical properties of the output radiation generated by a
degenerate parametric oscillator coupled to a squeezed vacuum
reservoir using the input-output relation introduced by Gardiner
and Collett \cite{9}. We also make a comparison with the results
of the cavity radiation. For both the cavity and output radiations
a maximum squeezing is occured at a critical point,
$\kappa=2\varepsilon$. The maximum squeezing for a cavity
radiation when the degenerate parametric oscillator is coupled to
a vacuum reservoir is found to be $50\%$, irrespective of the
cavity damping constant. The same result has been obtained by many
authors [1-4] using various approaches. However, we see from Eq.
\eqref{dp40} that the squeezing for the output radiation increases
with the damping constant at critical point. One can also clearly
see from Fig. 3 that the squeezing of the cavity radiation is
greater than that of the output radiation.

Coupling the degenerate parametric oscillator with the squeezed
vacuum reservoir is found to exponentially enhance the squeezing
of the output as well as the cavity radiation. Our calculation of
the squeezing spectrum for the output radiation indicates that a
complete quenching of the noise in one of the quadrature
components is possible at the critical point, even when the
degenerate parametric oscillator is coupled to the vacuum
reservoir. Since the squeezing spectrum corresponds to the
correlation of the quadrature operators at different times, we
note that the complete suppression of the noise could be achieved
if correlation at different times results squeezing and there is a
way of measuring the overall squeezing over sufficiently long
period of time. We also see from Fig. 4 that the width of the
squeezing spectrum decreases with the amplitude of the driving
radiation.

As the degree of  squeezing of the output and cavity radiations
increases with the amplitude of the coherent radiation and squeeze
parameter so does the mean photon number. This indicates that the
degenerate parametric oscillator produces not only a light with
high degree of squeezing but also a quite intense light at the
critical point. Moreover, even though only even number of photons
are produced by the degenerate parametric oscillator, there is a
finite probability for counting odd number of photons outside the
cavity. Furthermore, the power spectrum for the cavity and output
radiations is found to be picked at $\omega=0$. As clearly shown
in Fig. 5 the height of the power spectrum for the cavity
radiation is greater than that of the output radiation, and yet
have the same width. It is not difficult to see from Eqs.
\eqref{dp68} and \eqref{dp69} that the width of the power spectrum
is independent of $r$.


\begin{thebibliography}{1}
\bibitem{1} B. Daniel and K. Fesseha,
Opt. Commun. {\bf 151}, 384 (1998).
\bibitem{2} J. Anwar and M. S. Zubairy,
Phys. Rev. A {\bf45},
 1804 (1992).
\bibitem{3} K. Fesseha,
Opt. Commun. {\bf 156}, 145 (1998).
\bibitem{4} B. Teklu, Opt. Commun.
{\bf 261}, 310 (2006).
\bibitem{5}M. J. Collett and C. W. Gardiner,
 Phys. Rev. A {\bf30}, 1386 (1984).
\bibitem{6} L. A. Wu, M. Xiao, and H. J. Kimble,
 J. Opt. Soc. Am. B {\bf4}, 1465 (1987).
\bibitem{7} P. Gardiner, R. E. Slusher, B. Yurke, and A. Laporta,
 Phys. Rev. Lett. {\bf 59}, 2153 (1987).
\bibitem{8} J. Gea-Banacloche, Phys. Rev.
Lett. {\bf 59}, 543 (1987).
\bibitem{9} C. W. Gardiner and M. J. Collett,
Phys. Rev. A {\bf31}, 3761 (1985).
\bibitem{10} W. H. Louisell, {\it{Quantum statistical properties of radiation}} (Wiley, Newyork,
1973).
\bibitem{11}W. Chow, W. Koch, and M. Sargent III, {\it
Semiconductor-Laser Physics} (Springer-Verlag, Berlin, 1994).
\end{thebibliography}
\end{document}